\documentclass{article}
\usepackage{graphicx}
\usepackage{amsmath, amsthm, latexsym}
\usepackage{amssymb}
\begin{document}

\vspace*{3cm} \thispagestyle{empty}
\noindent \textbf{\Large Frame Dragging Anomalies for Rotating Bodies}\\

\noindent \textbf{\normalsize Peter Collas}\footnote{Department of Physics and Astronomy, California
State University, Northridge, Northridge, CA 91330-8268. Email: peter.collas@csun.edu.}
\textbf{\normalsize and David Klein}\footnote{Department of Mathematics, California State University,
Northridge, Northridge, CA 91330-8313. Email: david.klein@csun.edu.}\\

\vspace{4mm} \parbox{11cm}{\noindent{\small Examples of axially symmetric solutions to
Einstein's field equations are given that exhibit anomalous ``negative frame dragging'' in the sense
that zero angular momentum test particles acquire angular velocities in the opposite direction of
rotation from the source of the metric.}
\vspace{5mm}\\
\noindent {\small KEY WORDS:  Frame dragging; Kerr-Newman; Bonnor; van Stockum; 
Brill-Cohen spacetimes.}}\\\vspace{6cm}
\pagebreak

\setlength{\textwidth}{27pc}
\setlength{\textheight}{43pc}
\noindent \textbf{{\normalsize 1. INTRODUCTION}}\\

\noindent The prototype example of frame dragging arises in the Kerr metric. 
A test particle with zero angular momentum released from a nonrotating frame, far from
the source of the Kerr metric, accumulates nonzero angular velocity in the same
angular direction as the source of the metric, as the test particle plunges toward the
origin (in Boyer-Lindquist coordinates).  This ``dragging of inertial frames,'' or
frame dragging, is due to the influence of gravity alone, and has no counterpart in
Newtonian physics.

Frame dragging is a general relativistic feature, not only of the exterior
Kerr solution,  but of all solutions to the Einstein field equations 
associated with rotating sources.  In this paper we show that surprising frame dragging anomalies
can occur in certain situations.  We give examples of axially symmetric solutions to the field
equations in which zero angular momentum test particles, with respect to nonrotating coordinate
systems, acquire angular velocities in the opposite direction of rotation from the sources of the
metrics. We refer to this phenomenon as ``negative frame dragging.''

The mathematical considerations in this paper are straightforward, 
but from a physical point of view, negative frame dragging is counterintuitive
and intriguing.  The negative frame dragging in some of the models we consider is
associated with closed timelike curves due to singularities, and one might therefore expect to explain
the phenomenon entirely in terms of temporal anomalies (Bonnor [1], Kerr-Newman [2]). 
However, we also  show that negative frame dragging occurs relative to nonrotating,
inertial observers on the axes of symmetry of metrics that are completely free of causality
violations and singularities,  such as the low density, slowly rotating van Stockum dust cylinder [3],
(see also Tipler [4]), and the slowly rotating spherical shell of Brill and Cohen [5].  

In Section 2 we define frame dragging, and introduce notation.  In Section 3 we prove
the existence of negative frame dragging  for a model of a rotating dust cloud obtained by Bonnor [1]
and investigated by Steadman [6], and for the Kerr-Newman metric [2].  Section 4 contains a proof of
the existence of negative frame dragging for the low mass van Stockum dust cylinder [3], and Brill and
Cohen's slowly rotating spherical shell [5]. Our concluding remarks are in
Section 5.\\

\noindent \textbf{{\normalsize 2. FRAME DRAGGING}}\\

\indent A convenient way of writing the general stationary axisymmetric metric (vacuum or 
nonvacuum) is
\begin{equation}
ds^{2}=-F(dt)^{2}+L(d\phi)^{2}+2Mdtd\phi+H_{2}(dx^{2})^{2}+H_{3}(dx^{3})^{2}\;,
\end{equation}
where $F, L, M, H_{2}, H_{3}$ are functions of $x^{2}$ and $x^{3}$ only; consequently the 
canonical momenta $p_{t}$ and $p_{\phi}$ are conserved along geodesics.  From (1) 
we find that
\begin{eqnarray}
p_{t}&=&-F\dot{t}+M\dot{\phi}\equiv-E\,,\\
p_{\phi}&=&M\dot{t}+L\dot{\phi}\,.
\end{eqnarray}
The overdot stands for $d/d\tau$ 
for timelike particles and $d/d\lambda$ 
for lightlike particles, where $\tau$ denotes proper time, and $\lambda$ is an affine parameter.
$E$ and $p_{\phi}$ are the energy and angular momentum, respectively, of massless or massive 
particles.  We may then write, 
\begin{eqnarray}
\dot{t}&=&\frac{Mp_{\phi}+LE}{FL+M^{2}}\,,\\
\dot{\phi}&=&\frac{Fp_{\phi}-ME}{FL+M^{2}}\,.
\end{eqnarray}
\noindent Thus,
\begin{equation}
\frac{d\phi}{dt}=\frac{\dot{\phi}}{\dot{t}}=\frac{Fp_{\phi}-ME}{Mp_{\phi}+LE}\,.
\end{equation}
If we let $p_{\phi}=0$ in Eq. (6), we obtain the angular velocity $\omega$ of a zero angular 
momentum particle as measured by an observer for whom $t$ is the proper time.  This is the 
angular velocity of the frame dragging and it is given by,
\begin{equation}
\omega=-\frac{M}{L}\,.
\end{equation}\\

\noindent \textbf{{\normalsize 3. SINGULAR METRICS}}\\

\indent A solution to the field equations given by Bonnor in [1] describes
a cloud of rigidly rotating dust particles moving along
circular geodesics in hypersurfaces of $z=\mbox{constant}$. In contrast to the 
van Stockum dust cylinder considered in the next section, this metric has a singularity at $r=z=0$.
Bonnor's metric has the form of Eq.(1) where $F, L, M,$ and $H \equiv H_{2}=H_{3}$ are functions of
$x^{2}=r$ and $x^{3}=z$.  In comoving  (i.e., corotating)  coordinates these functions are given by
\begin{equation}
F=1\,,\;\;\;\;L=r^{2}-n^{2}\,,\;\;\;\;M=n\,\;\;\;\;H=e^{\mu}\,,
\end{equation}
\noindent where
\begin{equation}
n=\frac{2hr^{2}}{R^{3}}\,,\;\;\;\;\;\mu=\frac{h^{2}r^{2}(r^{2}-8z^{2})}{2R^{8}}\,,\;\;\;\;\;R^{2}=
r^{2}+z^{2}\,,
\end{equation}
and we have the coordinate condition
\begin{equation}
FL+M^{2}=r^{2}\,.
\end{equation}

\noindent The rotation parameter $h$ has dimensions of length
squared.  We assume without loss of generality, as in [6], that $h>0$.  The energy density $\rho$ is given by
\begin{equation}
8\pi \rho=\frac{4e^{-\mu}h^{2}(r^{2}+4z^{2})}{R^{8}}\,.\\
\end{equation}
\noindent As $R\rightarrow \infty$, $\rho$ approaches zero rapidly and the metric 
coefficients tend to Minkowski values. Moreover, all the Riemann
curvature tensor elements vanish at spatial infinity. Thus an observer at spatial infinity may be
regarded as  nonrotating, as in the case of the Kerr metric (in Boyer-Lindquist coordinates).

\indent Steadman [6] observed that null geodesics with angular momentum $p_{\phi}$ are
restricted to the region
$S_{B}$ given by
\begin{equation}
S_{B}=\{(t,\phi, r,z)|-p^{2}_{\phi}+2nEp_{\phi}+(r^{2}-n^{2})E^{2}\geq0\}\,.
\end{equation}
 
\noindent For the case where $p_{\phi}=0$  we let $S_{B}=S_{B0}$. Then 
$S_{B0}=\{(t,\phi, r,z)|L\geq 0\}$, and
$\partial S_{B0}=\{(t,\phi, r,z)|L=0\}$. The proof of the next proposition follows from 
direct calculation, using Eq. (7).\\

\noindent \textbf{Proposition 1}:  \textit{In Bonnor's metric,  $\omega
\rightarrow 0$ as either $r$ or $z$ go to $\infty$, $\omega <0$ everywhere in
$S_{B0}$, and $\omega\rightarrow -\infty$ on} $\partial S_{B0}$. \\ 
 
\indent Since $\omega \rightarrow 0$ as either $r$ or $z$ go to $\infty$, an observer at spatial
infinity observes a zero angular momentum test particle to be nonrotating (at infinity). The same
observer observes negative frame dragging at all finite $r$ and $z$ coordinate values in $S_{B0}$.  This
negative frame dragging is associated with temporal anomalies as we explain at the end
of this section.

\indent The Kerr-Newman metric [2] is a vacuum metric. It is a generalization of 
the Kerr metric that accounts for an electrical charge of the source. We write it below in
Boyer-Lindquist  coordinates.  Using the notation of Eq. (1) where
$F, L, M, H_{2}, H_{3}$ are now functions of $x^{2}=r$ and $x^{3}=\theta$, we have
\begin{equation}
F=\frac{\Delta-a^{2}\sin^{2}\theta}{\rho^{2}}\,,\;\;\;\;L=\frac{[(r^{2}+a^{2})^{2}-a^{2}\Delta\sin^{2}
\theta]\sin^{2}\theta}{\rho^{2}}\,,
\end{equation}
\begin{equation}
M=-\frac{a\sin^{2}\theta
(r^{2}+a^{2}-\Delta)}{\rho^{2}}\,,\;\;\;\;H_{r}=\frac{\rho^{2}}{\Delta}\,,\;\;\;\; 
H_{\theta}=\rho^{2}\,,
\end{equation}
where
\begin{equation}
\Delta=r^{2}+a^{2}+e^{2}-2mr\,,\;\;\;\;\mbox{and}\;\;\;\;\rho^{2}=r^{2}+a^{2}\cos^{2}\theta\,.
\end{equation}
In Eqs. (13)-(15) $m$ is associated with the mass of the source of the metric, $e$ is 
the electric charge, and the parameter $a$ is the angular
momentum per unit mass.  We note that the Kerr-Newman metric differs from the Kerr metric only in
the definition of $\Delta$.  For simplicity we consider only the case where
$a^{2}+e^{2}>m^{2}$.  In this case the Kerr-Newman metric has a naked (ring) singularity at
$\rho^{2}=0$.  There are no event horizons since under the above condition $\Delta>0$ for all $r$.
We also have that
\begin{equation}
FL+M^{2}=\Delta \sin^{2}\theta\,,
\end{equation}
thus $FL+M^{2}>0$ for $\theta\neq0$\,.\\
\indent The Kerr-Newman metric also has a ``forbidden region'' like the one found by Steadman [6]
for Bonnor's metric.  From Eqs. (14) and (15) we see that
\begin{equation}
H_{r}dr^{2}+H_{\theta}d\theta^{2}\geq 0\,,
\end{equation}
and since for null paths $ds^{2}=0$, it follows that
\begin{equation}
F\dot{t}^{2}-L\dot{\phi}^{2}-2M\dot{t}\dot{\phi}\geq 0\,.
\end{equation}
Now substituting Eqs (4) and (5) into Eq. (18), we obtain the inequality,
\begin{equation}
-Fp_{\phi}^{2}+LE^{2}+2Mp_{\phi}E\geq 0\,,\;\;\;\theta \neq 0\,.
\end{equation}
Null geodesics with angular momentum $p_{\phi}$ are restricted to a region $S_{KN}$ given by
\begin{equation}
S_{KN}=\{(t,\phi,r,\theta)|-Fp_{\phi}^{2}+LE^{2}+2Mp_{\phi}E\geq 0\}\,.
\end{equation}
For the case where $p_{\phi}=0$ we let $S_{KN}=S_{KN0}$. Then 
$S_{KN0}=\{(t,\phi,r,\theta)|L\geq 0\}$, and
$\partial S_{KN0}=\{(t,\phi,r,\theta)|L=0\}$. We then have the following proposition:\\

\noindent \textbf{Proposition 2}:  \textit{In the Kerr-Newman metric, $\omega$ vanishes at $\infty$ 
and at $r=e^{2}/(2m)$.  Furthermore $\omega\rightarrow -\infty$ on $\partial S_{KN0}$.}\\

\noindent\textbf{Proof.}  From Eqs. (7) and (13)-(15) we have that
\begin{equation}
\omega=-\frac{M}{L}=\frac{a(2mr-e^{2})}{B(r)}\,,
\end{equation}
where
\begin{eqnarray}
B(r)&=&(a^{2}+r^{2})^{2}-a^{2}\Delta\sin^{2}\theta\nonumber\\
    &=&r^{4}+a^{2}(2-\sin^{2}\theta)r^{2}+(2a^{2}m\sin^{2}\theta)r\nonumber\\
    & &-a^{2}\left[(a^{2}+e^{2})\sin^{2}\theta-a^{2}\right]\,.
\end{eqnarray}
It is clear that $\omega$ vanishes at $r=e^{2}/(2m)$ and at infinity.  We wish to show that 
$B(r)$ has one positive root and that this root cannot coincide with the root of the numerator.\\
\indent By Descartes' theorem $B(r)$ can have at most one positive root if
\begin{equation}
(a^{2}+e^{2})\sin^{2}\theta>a^{2}\,.
\end{equation}
Since the negative roots of $B(r)$ are the same as the positive roots of $B(-r)$, it follows
that, under condition (23), $B(r)$ can have at most one negative root.  Since
$B(0)<0$ if condition (23) holds, we see that indeed $B(r)$ has exactly one positive,
one negative, and two complex roots.  The positive root is the root of $L(r)$ that causes
$\omega\rightarrow -\infty$.  The proof is completed by observing that the equation $B(r=e^{2}/(2m))=0$
cannot be satisfied  for any real
$\theta$.

Referring to Eq. (1), consider a curve with fixed $(t, x^{2}, x^{3})$ coordinates, i.e.,
an integral curve of the $\phi$ coordinate. When $L>0$, this curve is a closed spacelike
curve of length given by $s^2=L(2\pi)^2$. However, when $L<0$,  it is a closed timelike curve, while 
when $L=0$, it is a closed null curve.  The last two cases are examples of causality violating paths.  
Thus, the forbidden regions for $p_{\phi}=0$ for the Bonnor and Kerr-Newman metrics coincide 
with the region described here where closed azimuthal timelike curves first appear. The sign of the 
metric coefficient $L$ determines the sign of the frame dragging, $\omega$, as well as a region  of
causality violations for integral curves of the $\phi$ coordinate. While negative
frame dragging may be correlated in this way with temporal anomalies for the Bonnor and Kerr-Newman
metrics, the metrics considered in Section 4 are free of causality violations. 
Yet, negative frame dragging occurs in nonrotating reference frames for those metrics.\\

\noindent \textbf{{\normalsize 4. NONSINGULAR METRICS}}\\

\indent The van Stockum solution [3] represents a rotating dust cylinder of infinite extent 
along the axis of symmetry ($z$-axis) but of finite radius.  There are three
vacuum exterior solutions that can be matched to the interior solution, depending on the
mass per unit length of the interior. Bonnor [7] labeled these exterior solutions: 
(I) the low mass case, (II) the null case, (III) the ultrarelativistic case. Tipler [4] (see also Steadman
[8]) showed that in case III, there exist causality violating paths in the
spacetime.  We focus on the low mass case, as it is the most physically
realistic of the three, and the most significant from the point of view of frame dragging.
We comment briefly on the frame dragging properties of the other two cases.

For the van Stockum metric, $H_{2}=H_{3}=H$ in Eq. (1), $x^{3}=z$, and the functions $F, L, M, H$ 
depend only on $x^{2}=r$.  The metric coefficients for the interior of the cylinder in comoving
coordinates, i.e., coordinates corotating with the dust particles, are given by
\begin{equation}
F=1\;,\;\;\;\;L=r^{2}(1-a^{2}r^{2})\;,\;\;\;\;M=ar^{2}\;,\;\;\;\;H=e^{-a^{2}r^{2}}\;
\end{equation}
In Eq. (24), $0\leq r \leq R$ for a constant $R$ that determines the radius of the cylinder, 
$a$ is the angular velocity of the dust particles, and the density $\rho$ is given by
$8\pi\rho=4a^{2}e^{a^{2}r^{2}}$.  The coordinate  condition
\begin{equation}
FL+M^{2}=r^{2}\,,
\end{equation}
holds for the interior as well as the exterior solutions below.  Furthermore since
$g=\det(g_{\mu\nu})=-(FL+M^{2})H^{2}=-r^{2}H^{2}<0$, the metric signature is $(-,+,+,+)$ 
for all $r>0$, provided $H\neq 0$; this is true in particular even if $L$ changes sign.\\
\indent The low mass vacuum exterior solution (Case I) is valid for  $0<aR<1/2$ and
$r\geq R$. The metric coefficients are
\begin{equation}
F=\frac{r\sinh(\epsilon-\theta)}{R\sinh\epsilon}\,,\;\;\;\;
L=\frac{Rr\sinh(3\epsilon+\theta)}{2\sinh 2\epsilon\cosh\epsilon}\,,
\end{equation}
\begin{equation}
M=\frac{r\sinh(\epsilon+\theta)}{\sinh 2\epsilon}\,,\;\;\;\;H=e^{-a^{2}R^{2}}
\left(\frac{R}{r}\right)^{2a^{2}R^{2}}\,,
\end{equation}
with
\begin{equation}
\epsilon=\tanh^{-1}(1-4a^{2}R^{2})^{1/2}\,,\;\;\;\;\theta=(\tanh\epsilon)\log\left(\frac{r}{R}
\right)\,.
\end{equation}

\noindent The metric is globally regular, and the algebraic invariants of the Riemann
tensor vanish as $r\rightarrow \infty$ (this is true also in Cases II and III provided $aR<1$).\\
\indent We consider noncomoving coordinates given  by the transformation
\begin{equation}
t=\bar{t}\;,\;\;\;\;\phi=\bar{\phi}-\Omega \bar{t}\;,\;\;\;\;r=\bar{r}\;,\;\;\;\;z=\bar{z}\;,
\end{equation}

\noindent where the barred coordinates are noncomoving.  In the barred coordinates, the metric
coefficients are:
\begin{equation}
\bar{F}=F+2\Omega
M-\Omega^{2}L\;,\;\;\;\bar{L}=L\;,\;\;\;\bar{M}=M-\Omega L\;,\;\;\;\bar{H}=H\;.
\end{equation}
\indent Among these barred coordinate systems, two values of $\Omega$ may be used to compute physically 
meaningful values of the angular velocity $\omega$ for frame dragging given by Eq. (7): 
$\Omega=a$ for an observer in a nonrotating inertial reference frame on the axis 
of symmetry, and $\Omega= \Omega_{c}$ (determined below) for an observer nonrotating
relative to ``the fixed stars.'' The choice $\Omega=a$ is determined by the Fermi-Walker equations.
A coordinate system satisfying the Fermi-Walker equations is rotation free (Walker [9], also Misner,
Thorne and Wheeler [10]), and it is therefore natural to study  frame dragging in such a coordinate
system. 

\indent By changing from polar to Cartesian coordinates, it is easy to
see that the spacetime may be extended to include the axis of symmetry ($r=0$), and 
the metric is Minkowskian there for any value of $\Omega$.  Furthermore, the reference frame of a
nonmoving observer with four-velocity,  $\vec{u} = (F^{-1/2},0,0,0)$ and orthonormal spatial frame
vectors in the $x,y,z$ directions satisfies the Fermi-Walker equations if and only if $\Omega=a$. 
Indeed, the fixed observer with this four-velocity lies on a geodesic, and the orthonormal
frame satisfies the parallel transport equations when $\Omega=a$. The calculations are 
straightforward. We note that in [3], van Stockum already argued that an observer on the axis 
with the above four-velocity is nonrotating if and only if $\Omega=a$, through a calculation
that involved taking limits as $r\rightarrow 0$ in cylindrical coordinates (our transformation, Eq.
(29), differs by a sign from the one that van Stockum used in ref. [3], p. 145).\\

\noindent \textbf{Proposition 3.} \textit{In the nonrotating, inertial reference 
frame of the low mass van Stockum cylinder corresponding to $\Omega=a$ described above, 
a zero angular momentum test particle experiences negative frame 
dragging at all points in the exterior and interior of the cylinder off of the axis of 
symmetry, that is, a zero angular momentum test particle with positive $r$ coordinate will 
accumulate an angular velocity in the direction opposite to the rotation of the cylinder. 
Furthermore, the angular velocity, $\bar{\omega}(r)$, given by }Eq. (7) \textit{for this coordinate
system, decreases monotonically to  the negative constant $a
-\left(2/R\right)e^{-2\epsilon}\cosh\epsilon\,$,  as} $r\rightarrow\infty$.\\
 
\noindent\textbf{Proof.} From Eqs. (7), (29), and (30),
$\bar{\omega}(r)\equiv-\bar{M}(r)/\bar{L}(r)= -M(r)/L(r)+a$. A simple calculation using Eq. 
(24) shows that $\bar{\omega}(r)$ is negative  whenever $0<r\leq R$. A second calculation using 
Eqs. (26) and (27) shows that $d\bar{\omega}/{dr}< 0$ for all $r \geq R$. 
It follows that $\bar{\omega}(r)<0$ for all $r>0$, and that $\bar{\omega}(r)$ is a decreasing function
of $r$. The limiting value
$\bar{\omega}(r\rightarrow\infty)=a-\left(2/R\right)e^{-2\epsilon}\cosh\epsilon\,$  follows directly
from Eqs. (26) through (30).\\
\indent Instead of $\Omega=a$, we may choose another value, $\Omega=\Omega_{c}$, in Eqs. 
(29) and (30) where $\Omega_{c}\equiv\Omega_{c}(a,R)$ is the ``critical value'' of $\Omega$ 
for which $\bar{\omega} (r\rightarrow \infty)=0$.  Such an $\Omega_{c}$ exists for Case I (as well as
Case II but not for Case III).  A short calculation shows that
\begin{equation}
\Omega_{c}=\left(\frac{2}{R}\right)e^{-2\epsilon}\cosh\epsilon\,,
\end{equation}

\noindent When $\Omega=\Omega_{c}$ it follows that $\bar{\omega}(r)>0$ for all
${r}$.  In this coordinate system $\bar{t}$ is the proper time of an observer at
$\bar{r}=r=0$ whose frame is nonrotating relative to the distant stars, i.e., 
relative to $r=\infty$. This observer does not observe negative frame dragging; but rather
the usual (positive) frame dragging in the angular direction of rotation
of the dust cylinder.\

We note that $g_{tt}=-F$ changes sign in the exterior
cases in the corotating coordinate systems.  However when we rotate the comoving
coordinates by $\Omega_{c}$, we have $g_{\bar{t}\bar{t}}<0$ for all
$\bar{r}$ in Case I. In the exterior Case II, the analogous critical value of $\Omega$  
results in $g_{\bar{t}\bar{t}}=0$ for all $\bar{r}$. Therefore in Case II $\partial/\partial
\bar{t}$ is a null vector.  Finally in Case III ($1/2<aR<1$) we have causality violating paths and
negative frame dragging that cannot be made positive.\\
\indent Two coordinate systems are determined by $\Omega=\Omega_{c}$ and 
$\Omega=a$, through Eqs. (29) and (30).  Nonrotating observers in frames
defined in terms of these coordinate systems observe completely different frame
dragging properties. In the first case, negative frame dragging occurs, while in the
second case it does not. Yet, both observers can claim to be nonrotating in physically
reasonable ways. In the first case, the observer is nonrotating in the sense that his
reference frame is nonrotating and is locally inertial, while the second observer
has the feature that the distant stars are fixed (i.e., nonrotating) in his frame. Van
Stockum in [3,11] already noted that these coordinate systems rotate relative
to each other, but it is a peculiar feature that zero angular momentum test particles
in one of these frames is dragged in the opposite angular direction from the motion of the
cylinder.\\
\indent Brill and Cohen [5] considered frame dragging and Machian
effects associated with a slowly rotating thin spherical shell of radius $r_{0}$ and mass $m$. 
They calculated the metric solution to the Einstein field equations to first order in the
angular velocity $a$  of the spherical shell. Their metric may be written in the form of Eq. (1) with 
$x^{2}=r$ $x^{3}=\theta$, and for $r>r_{0}$ the metric coefficients are given by 
\begin{equation}
H_{r} = \left(1+\frac{m}{2r}\right)^{4}\\,\;\;\;\;
H_{\theta}=r^{2}H_{r},
\,\,\,\,L=H_{\theta}\sin^{2}\theta,
\end{equation}
\begin{equation}
M=-L\omega (r)\,,\;\;\;\;F=\left(\frac{2r-m}{2r+m}\right)^{2}-Lw^{2}(r)\,,
\end{equation}

\noindent where
\begin{equation}
\omega(r) =\frac{4amr^{3}(m+2r_{0})^{5}(m-4r_{0})}{r_{0}^{3}(m+2r)^{6}(m-6r_{0})}\,\,\,\,\,\,\,
\mbox{for}\,\,\,r > r_{0}\\.
\end{equation}

\noindent We note that the function $\omega^{2}(r)$ in $F$ is not required in the lowest
order approximation.  \noindent When  $r<r_{0}$, $ H_{r}, H_{\theta}, M, L, F$, and $\omega$ each take
constant values determined by their respective formulas in (32), (33), and (34) evaluated at $r=r_{0}$,
so that, for example,
\begin{equation}
\omega(r_{0})=\frac{4am(m-4r_{0})}{(m+2r_{0})(m-6r_{0})}\,\,\,\,\,\,\,\mbox{for}\,\,\,r< r_{0}.
\end{equation}

\indent We see that $\omega$ vanishes for $m=4r_{0}$ and diverges for $m=6r_{0}$.  This
unphysical behavior is undoubtedly due to the approximations. Outside of an interval containing
$4r_{0}$ and $6r_{0}$, $\omega$ is monotonically increasing in $m$.  However, the shell collapses for $m\geqslant2r_{0}$, so we restrict attention to  $m<2r_{0}$.\\
\indent In the interior of the spherical shell, $\omega$ is constant and the metric can be made diagonal
by simple rotation, Eq. (29) with $\Omega=-\omega$. Therefore an observer on the axis
with a coordinate frame associated with this change of coordinates satisfies the Fermi-Walker
equations and is nonrotating. With the notation of Eq. (30), we calculate the frame
dragging $\bar{\omega}(r)$ for this observer as follows: $\bar{\omega}(r)\equiv-\bar{M}(r)/\bar{L}(r)=
-M(r)/L(r) -\omega(r_{0}) =\omega(r)-\omega(r_{0})$.  It is easy to see from Eqs. (34) and (35) 
that $\omega(r_{0})>0$ and $\omega(r)\rightarrow 0$ as $r\rightarrow \infty$.   Brill and Cohen give
plots of $\omega/a$ versus $r/r_{0}$ for various masses, and they show that $\omega$ is maximum in the
interior of the shell and decreases monotonically for $r>r_{0}$.  It follows that the nonrotating
inertial observer on the axis will observe negative frame dragging for all $r>r_{0}$. An extensive analysis of rotating, charged mass shells, including ``antidragging'' effects was given in [12]. \\

\vspace{30pt}

\noindent \textbf{{\normalsize 5. CONCLUDING REMARKS}}\\

\indent The existence of negative frame dragging depends on the qualitative behavior of the metric 
coefficients in Eq. (1). The examples considered in this note suggest that 
the phenomenon is fairly widespread among axially symmetric solutions to the field equations.\\

\noindent \textbf{ACKNOWLEDGMENT}\\

\noindent The authors wish to thank Professor John Lawrence and especially Herbert Pfister and Marcus King for an additional reference and useful comments.\\

\noindent \textbf{{\normalsize REFERENCES}}

\begin{enumerate}
\def\labelenumi{[\theenumi]}
\item Bonnor, W. B. (1977). \textit{J. Phys. A: Math. Gen.} \textbf{10}, 1673.
\item Newman, E. T., Couch, E., Chinnapared, K., Exton, A., Prakash, A., and Torrence, R.
(1965). \textit{J. Math. Phys.} \textbf{6}, 918.
\item van Stockum, W. J. (1937). \textit{Proc. R. Soc. Edin.} \textbf{57}, 135.
\item Tipler, F. J. (1974). \textit{Phys. Rev. D} \textbf{9}, 2203.
\item Brill, D. R., and Cohen, J. M. (1966). \textit{Phys. Rev.} \textbf{143}, 1011.
\item Steadman, B. R. (1999). \textit{Class. Quantum Grav.} \textbf{16}, 3685.
\item Bonnor, W. B. (1980). \textit{J. Phys. A: Math. Gen.} \textbf{13}, 2121.
\item Steadman, B. R. (2003). \textit{Gen. Rel. Grav.} \textbf{35}, 1721.
\item Walker, A. G. (1935). \textit{Proc. Edin. Math. Soc.} \textbf{4}, 170.
\item Misner, C. W., Thorne, K. S., and Wheeler, J. A. (1973). \textit{Gravitation}, W. H.
Freeman, San Francisco.
\item van Stockum, W. J. (1938). \textit{Proc. R. Irish Acad.} \textbf{44}, 109.
\item Pfister, H., King, M. (2002) \textit{Phys. Rev. D} \textbf{65}, 084033.
\end{enumerate}

\end{document}